\documentclass[prb,reprint,superscriptaddress,amsmath,amssymb,nolongbibliography,aps]{revtex4-2}

\usepackage{graphicx}
\usepackage{dcolumn}
\usepackage{bm}
\usepackage{textcomp} 
\usepackage{hyperref}
\usepackage[mathlines]{lineno}

\usepackage[normalem]{ulem}
\usepackage{xcolor}

\begin{document}

\preprint{APS/123-QED}

\title{Quantifying sample heating by ablation-plasma\\ x-rays in laser-driven shock experiments}

\author{R.F. Smith}
 \email{smith248@llnl.gov}
\affiliation{%
 Lawrence Livermore National Lab, Livermore, CA 94550, USA 
}
\author{I.K. OCampo}%
\affiliation{%
 Lawrence Livermore National Lab, Livermore, CA 94550, USA 
}%
\author{S. Singh}%
\affiliation{%
 Lawrence Livermore National Lab, Livermore, CA 94550, USA 
}%
\author{T.M. Hutchinson}%
\affiliation{%
 Lawrence Livermore National Lab, Livermore, CA 94550, USA 
}%
\author{T.E. Lockard}%
\affiliation{%
 Lawrence Livermore National Lab, Livermore, CA 94550, USA 
}%
\
\author{J.H. Eggert}%
\affiliation{%
 Lawrence Livermore National Lab, Livermore, CA 94550, USA 
}%
\author{C.P. McGuire}%
\affiliation{%
 Lawrence Livermore National Lab, Livermore, CA 94550, USA 
}%
\author{A.L. Coleman}%
\affiliation{%
 Lawrence Livermore National Lab, Livermore, CA 94550, USA 
}%
\author{J.M. McNaney}%
\affiliation{%
 Lawrence Livermore National Lab, Livermore, CA 94550, USA 
}%
\author{S. Pandolfi}%
\affiliation{%
Sorbonne Université, Muséum National d’Histoire Naturelle, UMR CNRS 7590, Institut de Minéralogie,
de Physique des Matériaux et de Cosmochimie, IMPMC, 75005 Paris, France
}%
\author{\\B. Nagler}%
\affiliation{%
Linac Coherent Light Source, SLAC National Accelerator Laboratory, Menlo Park, CA, USA 
}
\author{D. McGonegle}%
\affiliation{%
Atomic Weapons Establishment,
Aldermaston, Reading, RG7 4PR, United Kingdom
}
\author{H.J. Lee}%
\affiliation{%
Linac Coherent Light Source, SLAC National Accelerator Laboratory, Menlo Park, CA, USA 
}%
\author{D. Khaghani}%
\affiliation{%
Linac Coherent Light Source, SLAC National Accelerator Laboratory, Menlo Park, CA, USA 
}%
\author{N. Czapla}%
\affiliation{%
Linac Coherent Light Source, SLAC National Accelerator Laboratory, Menlo Park, CA, USA 
}%
\author{A. Krygier}%
\affiliation{%
 Lawrence Livermore National Lab, Livermore, CA 94550, USA 
}%

\date{\today}

\begin{abstract}
\noindent
Sample preheat due to absorption of x-rays emitted by an ablation plasma has long been a source of uncertainty in laser-driven dynamic compression experiments.~This unknown heating complicates interpretation of experiments designed to reach well-defined thermodynamic states, such as the principal Hugoniot, particularly because the pressure drive and sample heating occur simultaneously and are difficult to disentangle.~Here, we quantify sample heating from x-rays using a unique experimental design in which the sample is heated prior to the arrival of the pressure wave. Using \emph{in situ} x-ray diffraction, we relate lattice expansion to temperature within a 1-µm-thick Au sample for commonly used 25-$\mu$m and 75-$\mu$m-thick Kapton-B (polyimide) ablators over a range of laser irradiances. Heating, sufficient to melt Au ($\Delta$T$\sim$1050~K), is observed for 25-µm-thick polyimide ablators, which is significantly enhanced by the addition of a 0.2-µm Al flash coating. In contrast, heating is strongly suppressed for 75-µm-thick, non-Al-coated Kapton-B, with no measurable heating observed in 2.5-µm-thick Ti or 1-µm-thick Fe samples at laser irradiances up to $1.2\times10^{13}$~W/cm$^{2}$.

\end{abstract}

\maketitle


\section{Introduction\label{sec:Introduction}}
\noindent
Compressibility \cite{Knudson:PRL2009, Smith2014, Fratanduono:SCI2021}, the location of phase boundaries \cite{dewaele2018, sharma2019, briggs2019, Li1987, coleman2019}, the magnitude of a material’s strength \cite{VOGLER:2006812, Brown:JAP2014, Krygier:PRL2019}, and deformation dynamics \cite{Smith2011} are all sensitive to sample temperature. In dynamic compression experiments, the thermal state depends on the compression history and on intrinsic heating from plastic work and phase transformations. Surrounding layers can further alter the temperature through wave reverberations and, in some cases, thermal conduction \cite{Sio:NCOMM2023}; the probe \cite{Krygier:POP2022} and drive \cite{Smith:POP2007, Krygier:PRB2026} sources can also heat the sample directly through x-ray absorption. A recent study by Singh et al. details many of these effects in the context of identifying shock melt \cite{Singh:PRB2026}. This work focuses on heating by x-rays emitted by the laser-produced ablation plasma, which are of particular concern because the source is close to the sample, the heating is concurrent with the drive, and the effects are difficult to measure or isolate.


High-pressure shock-compression experiments on high-power laser facilities typically employ a low atomic number material as an ablator \cite{Wicks2024,briggs2019,gorman2020,wark2025,sharma2019, coleman2019}. Laser irradiation at $10^{12}$--$10^{13}$~W/cm$^{2}$ rapidly heats the ablator, producing an expanding plasma 
\cite{schaeffer2016}.~Polyimide, commonly supplied from DuPont as Kapton-B/CB or Kapton-HN, is widely used due to its high ablation velocity, relative transparency to x-rays, and availability in thicknesses suitable for laser experiments.

Laser pulse shaping enables temporally-steady shock compression,  
with increasing laser power generating higher pressures. With current long-pulse capabilities at the Matter in Extreme Conditions (MEC) endstation of the Linac Coherent Light Source (LCLS), laser energies of up to 115~J delivered in 10~ns enable irradiances as high as $1.2\times10^{13}$~W/cm$^{2}$ within a $\sim$240-µm  diameter focal spot~\cite{Nagler2015}. At the highest irradiances this is sufficient to generate hundreds of GPa within a sample of interest. Several other laser facilities around the world have comparable capabilities \cite{gorman2022, mason2018, kim2019, inubushi2020, wang2019, guardalben2011, soures1996}. 

Laser-driven shock-compression experiments are commonly designed to access states along a sample's principal Hugoniot, and \emph{in situ} x-ray diffraction (XRD) is routinely used to diagnose the resulting material state \cite{Wicks2024,briggs2019,gorman2020,wark2025,sharma2019}. The principal Hugoniot is a particularly valuable target because it defines a reproducible thermodynamic path that is strongly constrained by the Rankine--Hugoniot relations and can be directly compared with existing shock data and equation-of-state (EOS) calculations. Interpreting experiments in this way requires the assumption that the sample undergoes shock compression without significant additional heating. Because samples are typically positioned only $\sim$25--75~\textmu m from the ablation plasma, x-ray preheat may violate this assumption and raise the sample temperature above the principal Hugoniot.


\begin{figure*}[!]
\begin{center}
\includegraphics[width=1\textwidth]{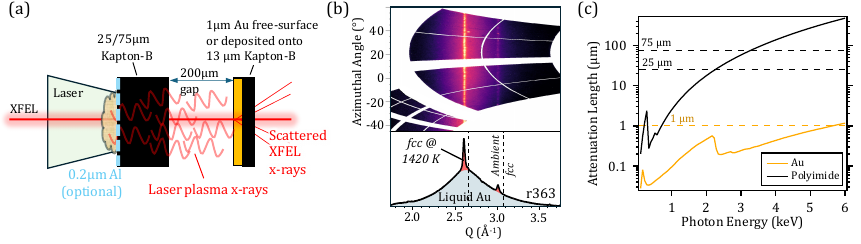}
\end{center}
\caption{\textbf{Experimental determination of heating from laser plasma x-rays. (a)} The experimental setup employs a 200-µm gap to temporally separate heating from laser plasma x-rays from subsequent compression wave heating.  \textbf{(b)} For shot r363, the level of heating was sufficient to partially melt the 1-µm-thick Au sample (Al-coated 25-µm Kapton-B, I$_{Laser}$$\sim$6$\times$10$^{12}$ W/cm$^2$). Lattice expansion of the unmelted face-centered cubic (fcc) crystal structure was determined to be consistent with a temperature of 1420 K ($\Delta$T = 1122 K). 
\textbf{(c)} 
X-ray attenuation length for polyimide and Au \cite{henke1993}.
}
\label{fig:Xray_heating}
\end{figure*}

In this work, we quantify x-ray--induced sample heating using XRD to measure lattice expansion, which is directly related to temperature ($T$) through established thermal expansion coefficients \cite{pamato2018}. 
These measurements use an ablator-gap-sample configuration that allows the sample to be probed after x-ray heating but before arrival of the pressure wave.~In this state, changes in lattice spacing can be attributed to thermal expansion alone, providing a direct measure of x-ray--induced sample heating.
~Measurements are performed primarily on 1-\textmu m-thick Au samples employing either 25- or 75-\textmu m-thick Kapton-B ablators, with and without a 0.2-\textmu m-thick Al flash coating applied to the laser-facing surface. Such coatings are commonly used to suppress laser shine-through \cite{sharma2019,briggs2019,Hodge2022,Madhavi2023,Coleman2022}. Au is chosen as it is highly absorbing to x-rays and, for the expected x-ray spectrum, the 1-µm thickness is efficiently heated.

We find that x-ray heating increases with both laser irradiance and time after laser onset, with the strongest heating observed for 25-\textmu m-thick Kapton-B ablators. Heating is further enhanced by the presence of an Al coating, with melting in Au ($\Delta$T$\sim1050$~K) observed for laser irradiances of 0.5 and 1.2$\times$10$^{13}$ W/cm$^{2}$, for 25-µm and 75-µm-thick Kapton-B, respectively. Minimal heating is observed for uncoated 75-µm-thick Kapton-B ablators.

\section{Experimental Setup}

\noindent
A schematic of the experimental setup and target design for experiments on MEC-LCLS is shown in Fig.~\ref{fig:Xray_heating}(a). The target consisted of either a 25-\textmu m or 75-\textmu m thick Kapton-B ablator, a 200-\textmu m gap, and a 1-\textmu m-thick Au foil, which was either free-standing or directly deposited onto a 13-\textmu m thick Kapton-B layer. For some samples, a 0.2-\textmu m thick Al layer was vapor deposited onto the laser-facing surface of the Kapton-B ablator. 
The gap between the Kapton-B ablator and the Au sample delays the arrival of the pressure wave until after the laser pulse ends. As a result, x-ray–driven heating and its associated thermal expansion occur before shock compression, allowing the expansion to be isolated and directly measured using x-ray diffraction.

\begin{figure}[t!]
\begin{center}
\includegraphics[width=1\columnwidth]{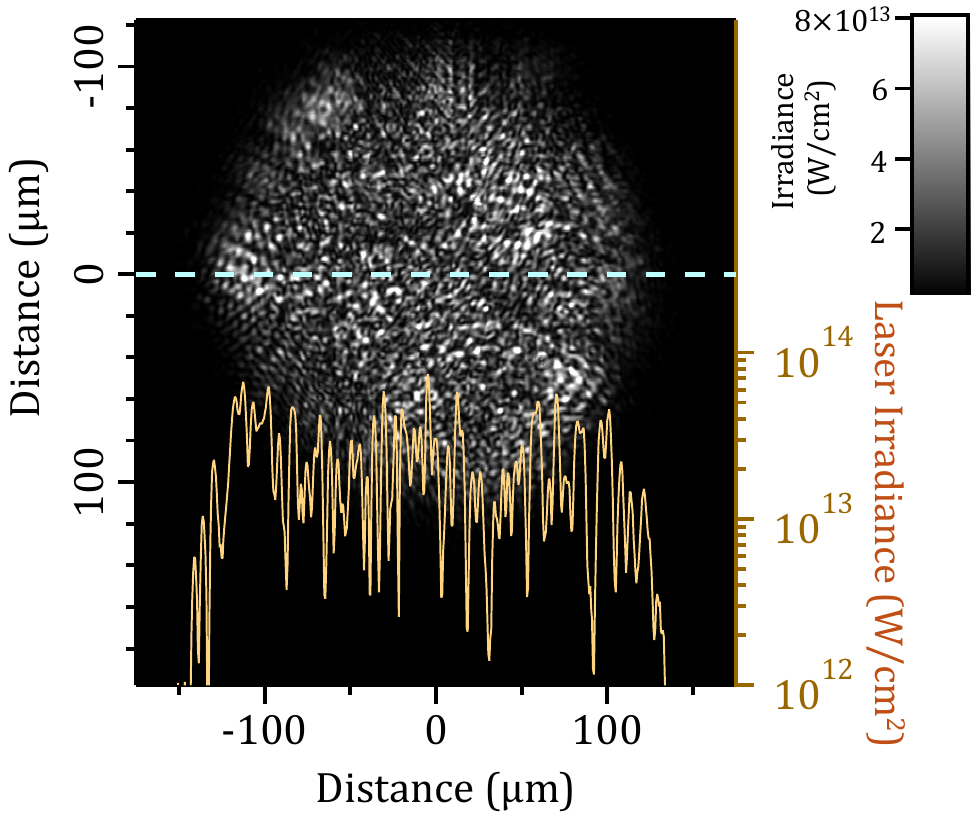}
\end{center}
\caption{\textbf{Time-integrated equivalent-target-plane laser focal spot at the 527-nm MEC laser.} The overall intensity envelope within the CPP-produced
laser spot is modulated by a high-frequency speckle. After accounting for the integrated laser energy and laser pulse duration (here assumed to be 100 J over 10 ns), a lineout of laser irradiance (brown trace) integrated over
a 0.5 µm wide region at the target plane (blue dashed horizontal line) is calculated. While these conditions give an average irradiance of 1.1$\times$10$^{13}$ W/cm$^2$, we observe local irradiance variations of 0.15$\rightarrow$7.5$\times$10$^{13}$ W/cm$^2$.
}
\label{fig:MEC_Intensity}
\end{figure}

The front surface of the Kapton-B ablator was positioned at the focal plane of four laser beams delivering a combined energy of up to 115 J at a wavelength of 527 nm in a 10-ns flat-top pulse. As laser focal spots are inherently non-planar, we employ continuous phase plates (CPPs), inserted into the beam path, to ensure spatially averaged uniform irradiance that is stable in time, ensuring high reproducibility of drive performance \cite{marozas2006, lin1996}. The laser focal spot is characterized by a reproducible irradiance envelope modulated by fine speckle (Fig. \ref{fig:MEC_Intensity}). At MEC, the minimum speckle size is estimated to be $\sim$4.2 µm \cite{gorman2022}. While the average irradiance is reported throughout this report, we note that there are localized regions of high irradiance that can result in enhanced x-ray emission.

Previous experiments have determined that, at MEC, 50\% of the laser energy is contained within a 240-µm diameter focal spot~\cite{Pereira2025}.~Absorption of the laser energy by the Kapton-B ablator produces a rapidly expanding ablation plasma, whose momentum drives a shock wave into the target assembly with a duration matched to the laser pulse.~A line-imaging velocity interferometer (VISAR)~\cite{Celliers2023} was used to determine the arrival time of the compression wave at the Au sample, with two VISAR channels employed with  velocity sensitivities of 0.6645 and 0.498 km s$^{-1}$ fringe shift$^{-1}$, respectively.~All temperature measurements were performed prior to shock arrival, with the majority of data taken at an XFEL probe time 10 ns after laser onset.

The 60-300-fs SASE (self-amplified spontaneous emission) output of the LCLS x-ray free-electron laser at a photon energy of $\sim$9.61~keV was incident on the target in a $\sim$30-\textmu m diameter spot centered on the laser drive region. The relative pointing and timing of the x-ray probe with respect to the laser-driven shock were known to within a few microns and $<100$~ps~\cite{Rastogi2022}. X-rays scattered from the heated Au sample were recorded in transmission geometry using multiple large-area ePix detectors~\cite{Dragone2014}. The detector positions in diffraction angle ($2\theta$) and azimuthal angle ($\phi$) were calibrated using diffraction patterns from an ambient-pressure CeO$_2$ and LaB$_6$ standards~\cite{Tracy2019}.

While Kapton is primarily polyimide plastic (C$_{22}$H$_{10}$N$_2$O$_5$), we conducted an analysis to determine the extent of trace elements within Kapton-B, and another commonly used variety (Kapton-HN), the presence of which could affect the absorption characteristics plotted in Fig. \ref{fig:Xray_heating}(c). To characterize trace metals in the Kapton films, 75-µm-thick Kapton-B and 50-µm-thick Kapton-HN samples were processed by direct microwave digestion and subsequently analyzed by inductively coupled plasma mass spectrometry (ICP-MS). Titanium was detected in both samples, at concentrations of 1.51 ppm in Kapton-B and 1.44 ppm in Kapton-HN. Aluminum and copper were additionally detected in Kapton-HN at 2.41 and 2.05 ppm, respectively, but were not detected in Kapton-B (Table \ref{Table:Concentration}). Iron, lead, manganese, nickel, and zinc were also tested for but were not detected in either sample. The detection limits for this digestion method were in the parts-per-million range; therefore, nondetection does not exclude the presence of these elements at lower concentrations.

\begin{table}[b!]
\begin{center}
\includegraphics[width=1\columnwidth]{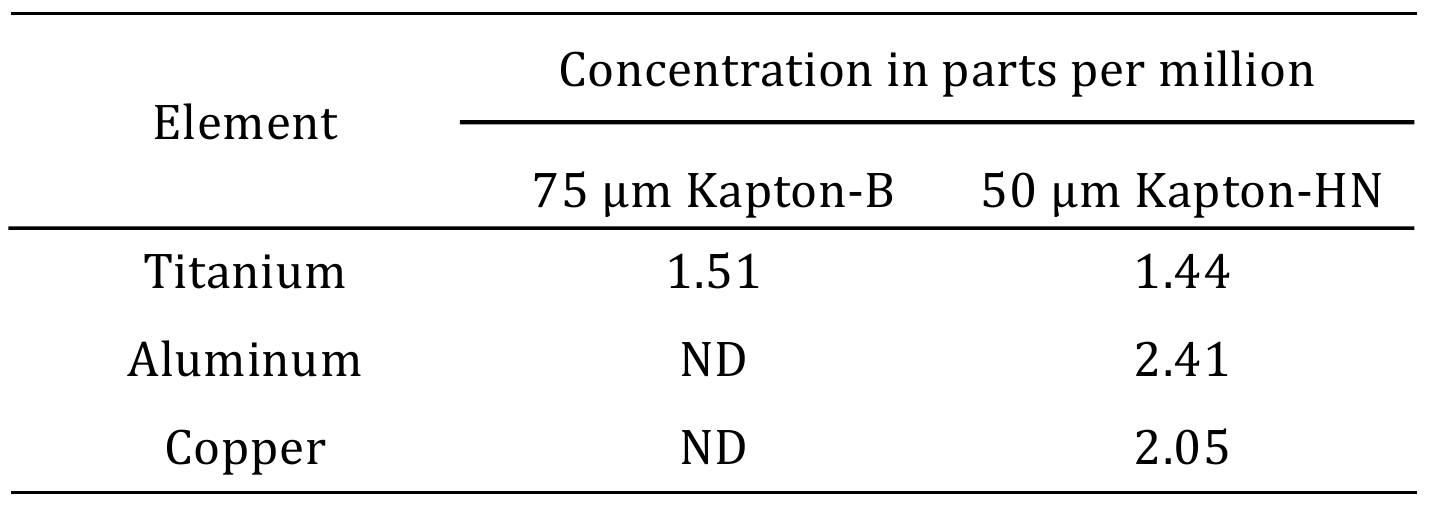}
\end{center}
\caption{\textbf{Concentration table. ND = Not detected.}  
}
\label{Table:Concentration}
\end{table}


The x-ray attenuation length from 0.1--6~keV for both polyimide and Au is plotted in Fig.~\ref{fig:Xray_heating}(c).~The attenuation length ($\alpha$) is defined as the depth into the material, measured along the surface normal, at which the x-ray intensity falls to $1/e$ (36.8\%) of its value at the surface \cite{henke1993}.~For a 1-$\mu$m-thick Au layer, this length is less than the sample thickness for all photon energies up to 6~keV, indicating efficient energy deposition. The x-ray absorption as a function of depth is
$-\frac{dI}{dx} = \frac{I_0}{\alpha} e^{-\mathrm{x}/\alpha}$, where I$_0$ is the incident x-ray flux at an x-ray energy E, $\mathrm{x}$ is the distance through the Au sample where $x$=0 is the plasma facing surface. Energy deposition and heating are therefore strongest on the plasma-facing surface. For the majority of shots, heating accumulates over 10 ns before x-ray diffraction provides a measure of the volume-integrated lattice expansion. Gold's ambient thermal diffusion length scale for 10 ns duration is $\sim1.1$~$\mu$m so we expect temperature gradients imposed by the strong absorption to be somewhat equilibrated within the 1-µm-thick Au sample.

\subsection{Determination of Temperature Rise}

To determine the temperature increase of the gold sample arising from plasma-induced x-ray emission during the laser drive, changes in the lattice parameter were extracted from x-ray diffraction measurements and compared against a calibrated thermal expansion model. The detector geometry was calibrated using the HEXRD software package \cite{HEXRD_github} with the LaB$_6$ and CeO$_2$ powder standards.~Following calibration, diffraction images from both ambient preshots and driven shots were azimuthally averaged to produce one-dimensional intensity profiles. Because the gold sample was spatially offset from the calibration standards by the combined thickness of the ablator and the ablator–sample gap, a geometric correction was applied to account for the resulting change in effective detector distance prior to data analysis.~This correction was applied consistently to all ambient and on-shot measurements. Lattice parameters were extracted using the Le Bail method and Gaussian peak fitting. 

Ambient lattice parameters were measured immediately prior to each driven shot. After each ambient measurement, the target was translated by approximately 100-$\mu$m before diffraction from the x-ray–heated sample was recorded, mitigating potential sample damage by the intense x-ray pulse, including x-ray–induced annealing, melting, and defect generation. Concurrent heating or damage by the probe x-rays is not expected to influence our lattice expansion measurements because the gold ions cannot respond on the $\sim$ 100~fs timescale of the probe x-ray pulse.

The temperature rise associated with absorption of x-rays emitted from the ablation plasma was inferred by comparing the measured on-shot lattice parameter to the expected thermal expansion of gold using the third-order Grüneisen thermal expansion model reported by Pamato \emph{et al.} \cite{pamato2018}. In this formulation, the temperature dependence of the unit-cell volume is determined by integrating the Grüneisen parameter and the isothermal bulk modulus, both of which are expressed as functions of volume. The refined lattice parameter was used to determine the unit-cell volume increase relative to ambient conditions. This lattice expansion was then related to sample temperature. The inferred temperatures fall well within the range over which the Pamato model has been validated \cite{pamato2018}.
This analysis assumes that the observed lattice expansion is dominated by thermal effects. While thermally induced stress waves may, in principle, contribute to the measured response, no evidence of such effects is observed in the VISAR measurements. Furthermore, the x-ray diffraction probe occurs prior to the shock traversing the ablator–sample gap and reaching the gold sample. 


An example diffraction pattern is shown in Fig.~\ref{fig:Xray_heating}(b), with an azimuthally averaged lineout plotted as a function of momentum transfer $Q$ [Q = 4$\pi$sin($\theta$)/$\lambda$, where $\lambda$ is the x-ray wavelength and $\theta$ is the Bragg scattering angle]. Here, we observe coexistence of solid and liquid scattering from the 1-$\mu$m-thick Au layer. The shift of the fcc peaks relative to the ambient-temperature reference is consistent with a temperature of 1420 $\pm$ 47 K ($\Delta T = 1122$~K).


\subsection{Temperature Uncertainty Analysis}

Uncertainties in the inferred temperature include: (1) detector calibration uncertainty, quantified by the offset between the refined LaB$_6$ lattice parameter and its accepted reference value; (2) uncertainty in the geometric position correction arising from finite precision in the ablator thickness and ablator–sample gap; (3) uncertainty associated with the centroid of the XFEL self-amplified spontaneous emission (SASE) spectrum, determined by fitting a pseudo-Voigt function to each individual spectrum and taking the standard deviation of the resulting centroids; (4) the standard deviation of the measured ambient gold lattice parameters; (5) the offset between the mean ambient lattice parameter and that predicted by the Pamato model at room temperature \cite{pamato2018}; and (6) the average Le Bail refinement uncertainty \cite{Lebail1995}.

The standard deviation of the SASE centroid was approximately 6 eV, which is small relative to the total bandwidth of the incident spectrum ($\sim$20-40 eV). The mean centroid was therefore used for lattice refinement, with the associated variability explicitly propagated into the temperature uncertainty. Each contribution was converted to an equivalent uncertainty in lattice parameter relative to the ambient Au lattice parameter of 4.0786 Å and subsequently propagated through the thermal expansion model to obtain a corresponding temperature uncertainty. The individual terms were combined in quadrature to yield the total temperature uncertainty reported for each shot. This approach provides an estimate of the uncertainty in the inferred temperature rise and captures both experimental and model-dependent contributions.

\subsection{Temperature gradients}

For the majority of driven shots, the diffraction peaks remain symmetric despite measurable lattice expansion, suggesting that temperature gradients along the beam direction are minimal and that the extracted lattice parameter represents a reasonable volume-averaged temperature.
However, in a subset of measurements, slight diffraction peak asymmetry is observed, which reflects a range of  temperature and density states  throughout the  1-µm sample thickness. 
This asymmetry is most evident for the higher angle peaks (see Figs. \ref{fig:Peak_Asymmetry} and \ref{fig:Peak_Asymmetry_75}). For these shots there is an additional error in determining the lattice parameter when using the Le Bail Gaussian peak fitting approach.~We have determined that the uncertainty from that Gaussian lineshape assumption within the Le Bail fitting can result in a $\sim$1$\%$ uncertainty in volume and an additional $\sim$100-K systematic uncertainty in the average temperature determination.  

For shots where a mixed solid–liquid phase assemblage is observed, the peak asymmetry is minimal because the crystalline diffraction originates from the colder back portion of the Au layer, while the liquid scattering originates from the hotter front portion of the Au layer (e.g., Fig. \ref{fig:Peak_Asymmetry}).~For these shots, the determined foil temperature may be a lower limit.

Indeed, we observe a characteristic change in the nature of the diffraction peak asymmetry for temperatures below melt (low diffraction angle tail) to temperatures where a mixed solid+liquid phase is observed (high diffraction angle tail) (Figs. \ref{fig:Peak_Asymmetry} and \ref{fig:Peak_Asymmetry_75}).~This we interpret as a change in the functional form of the temperature-depth profile.~Upon melting, some portion of the absorbed x-ray energy is used to break fcc-Au bonds (latent heat of fusion) rather than to raise the local Au temperature. We speculate that this may give rise to the change in the nature of diffraction peak asymmetry. However, a detailed analysis of this dependency is beyond the scope of this work. 

 \begin{figure}[!t]
\begin{center}
\includegraphics[width=1\columnwidth]{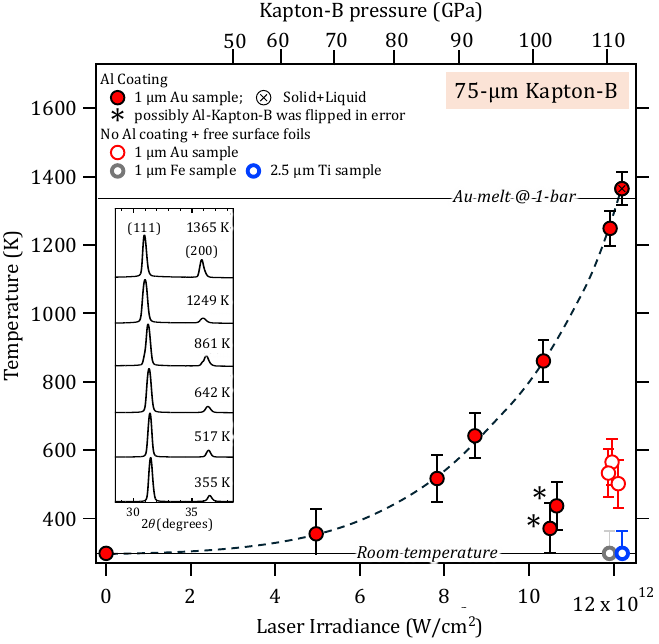}
\end{center}
\caption{\textbf{75-µm-thick Kapton-B ablators.} Filled red circular symbols represent the Au temperature in targets which employ an Al-coated 75 µm-thick Kapton-B ablator. All data were taken after 10-ns (laser pulse duration). Inset, diffraction lineouts for high-$T$ data along dotted black curve which shows clear lattice expansion (shift to lower 2$\theta$) as a function of increasing laser irradiance. The open red circles represent the same samples \underline{without} the Al coating. The gray and blue open circles represent the temperature from 1 µm Fe and 2.5 µm Ti, respectively (no measurable heating). The top axis reflects the expected pressure in the Kapton-B for a given laser irradiance~\cite{Pereira2025}.
}
\label{fig:75um_Kapton}
\end{figure}

\begin{figure}[t!]
\begin{center}
\includegraphics[width=1\columnwidth]{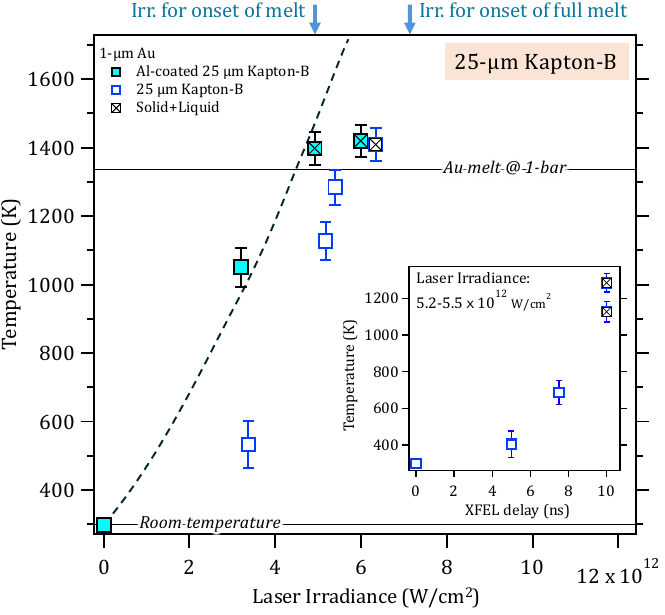}
\end{center}
\caption{\textbf{25-µm-thick Kapton-B ablators.} Data from Al-coated Kapton-B is shown by the filled cyan symbols. Data from the Kapton-B without Al is shown by the open symbols.~All data was taken after 10-ns. The arrows on the top axis indicate the first observation of solid-liquid coexistence (e.g., Fig. \ref{fig:Xray_heating}(b)) and complete melting. The inset figure shows the increase in temperature as a function of time for an approximately constant laser irradiance 
(see also Fig. \ref{fig:Peak_Asymmetry}). 
}
\label{fig:25um_Kapton}
\end{figure}


\section{Results}

\subsection{75-µm ablators}

In Fig.~\ref{fig:75um_Kapton}, we show the Au temperature as a function of laser irradiance for 75-$\mu$m-thick Kapton-B targets with an Al coating (filled symbols) and without an Al coating (open symbols). For all data in the main figure, an XFEL probe delay of 10~ns was used. In the absence of an Al coating, we observe a temperature rise of approximately 235~K in the Au foil at the highest laser irradiance investigated. We repeated this measurement by replacing the Au with either a 2.5-$\mu$m-thick Ti foil or a 1-$\mu$m-thick Fe foil; in both cases, no measurable lattice expansion was observed.

For 0.2-µm Al-coated Kapton-B ablators we observed stronger heating of the 1-µm Au foil as a function of increasing laser irradiance.~There are two noticeable outlier points to this trend, as indicated by the asterisk. The temperatures measured for these points are consistent with those obtained using uncoated Kapton-B targets. We therefore suspect that for these shots the Al-coated Kapton-B foil was inadvertently flipped during target assembly. In this scenario the Al would not be directly ablated by the laser drive.~At the highest laser irradiance the inferred Au temperature is consistent with the reported ambient pressure Au melt temperature \cite{weck2020}, although no clear observation of liquid is observed. At these laser irradiances the Au temperature difference between the Al-coated and non-Al-coated 75-µm-thick Kapton-B ablators is $\sim$800-K.



\bigskip
\subsection{25-µm ablators}

In Fig.~\ref{fig:25um_Kapton}, we show the Au temperature as a function of laser irradiance for 25-$\mu$m-thick Kapton-B targets with an Al coating (filled symbols) and without an Al coating (open symbols).~The Au foil temperatures are notably higher than those measured for the 75-$\mu$m-thick Kapton-B ablators, with the Au melt line reached at less than half the laser irradiance.

The fcc-Au temperature difference between the Al-coated and non-Al-coated 25-µm-thick Kapton-B ablators is $\sim$520~K for heating below melting ($I_{Laser}$$\sim$3.3$\times$10$^{12}$ W/cm$^2$). For higher laser irradiances this temperature difference is reduced as temperatures get pinned close to the expected 1-bar melt line \cite{weck2020}. This behavior is consistent with the effects of latent heat during melting, where a portion of the absorbed x-ray energy is used to break atomic bonds and form liquid rather than increase the bulk temperature.~Solid-liquid coexistence is first observed at $I_{Laser}$$\sim$5$\times$10$^{12}$ W/cm$^2$ (Fig.~\ref{fig:Xray_heating}(b)), with full melt at 7.1$\times$10$^{12}$ W/cm$^2$.

Figure \ref{fig:25um_Kapton} inset shows the temperature for 1-µm-thick Au as a function of time after the onset of laser illumination, for an average laser irradiance of $\sim$5.2-5.5$\times$10$^{12}$ W/cm$^2$. The temperature exhibits a non-linear dependency, with the strongest accumulation of heating from 5$\rightarrow$10 ns. We speculate that this dependency may relate to the evolution of volume and temperature of the expanding plasma.



\section{Discussion}


The use of 25-µm-thick Kapton-B ablators, both with and without an Al coating, results in considerably stronger sample heating (Fig.~\ref{fig:25um_Kapton}). For example, at a laser irradiance of $5\times10^{12}$~W/cm$^{2}$, Al-coated Kapton-B ablators with thicknesses of 25 and 75-µm yield measured Au temperatures of $1398~$K ($\Delta$T=$1100~$K) and $352~$K ($\Delta$T=$54~$K), respectively. This enhanced heating arises from the reduced x-ray absorption of the 25-µm-thick Kapton-B foil (Fig.~\ref{fig:Xray_heating}(c)). Even in the absence of an Al coating, 25-µm-thick Kapton-B ablators result in Au melting at laser irradiances exceeding $\sim$6$\times10^{12}$~W/cm$^{2}$. 


We note that the temperatures shown in Figs.~\ref{fig:75um_Kapton} and \ref{fig:25um_Kapton} are expected to be higher in conventional experiments with no ablator--sample gap \cite{sharma2019, briggs2019}. In the present geometry, the 200~$\mu$m gap reduces the fraction of the emitted x-rays that reaches the probed region. Removing the gap therefore increases the energy the sample absorbs, by a factor of 2.4--3.4 for the representative conditions discussed in Sec.~\ref{sec:gap_correction}. This factor is smaller than the geometric gain alone because, near contact, more of the intercepted photons arrive obliquely and are more strongly filtered by the Kapton.

\subsection{Gap Correction Estimation}\label{sec:gap_correction}

Our experiments placed a vacuum gap between the ablator and the probe layer to separate heating by x-rays from heating and compression by the associated high-pressure drive. The broader goal, however, is to understand this heating source in conventional shock-compression experiments, which have no such gap. There, the sample sits much closer to the x-ray emission region and absorbs correspondingly more energy.~Applying our measurements to that configuration requires accounting for this geometric difference, so we assess how the energy deposited in the sample scales with the ablator--sample gap. This assessment rests on an assumed emitted spectrum and angular distribution that our measurements do not directly constrain. The resulting corrections should therefore be read as estimates conditional on those assumptions, not as calibrated predictions.

\begin{figure}[!t]
\begin{center}
\includegraphics[width=1\columnwidth]{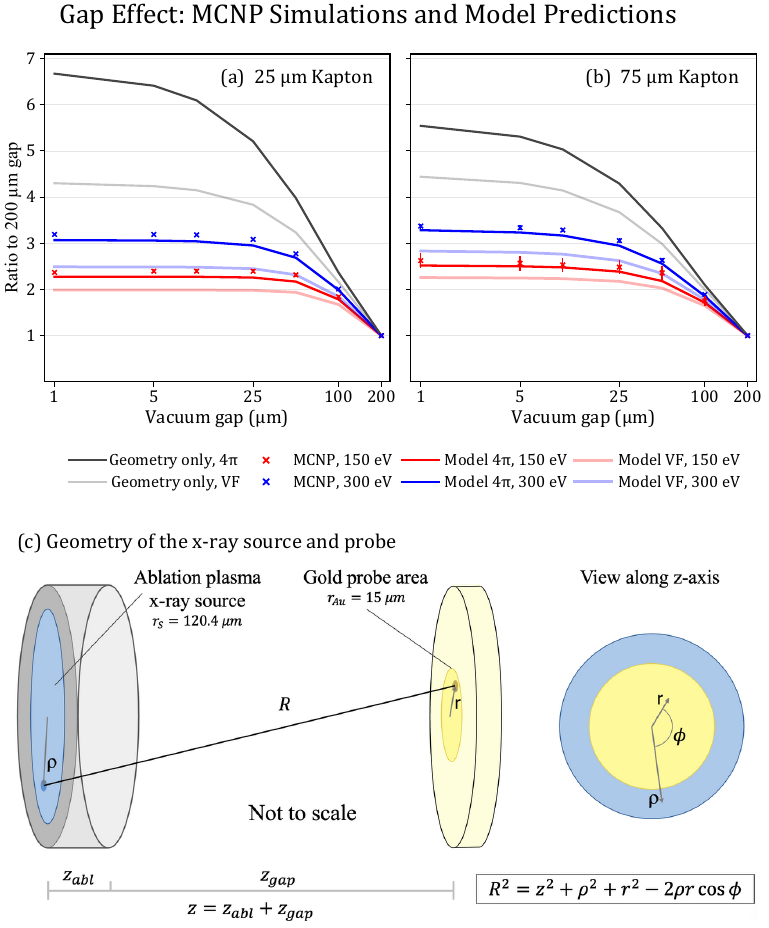}
\end{center}
\caption{\textbf{Gap and source-size dependence of Au energy deposition.} Panels (a,b) show the deposited-energy ratio relative to a 200 µm vacuum gap as the gap varies for a 240.8 µm-diameter x-ray source. Panel (a) uses 25 µm Kapton and panel (b) uses 75 µm Kapton; panel (c) sketches the source, ablator, and 30 µm-diameter Au probe region. The Au layer is 1 µm thick. Red and blue crosses are MCNP F6 photon-heating ratios for source spectral temperatures $kT=150$ and $300$ eV, respectively; error bars show MCNP statistical uncertainty and are mostly smaller than the markers. Dark red and blue solid lines are analytical absorbed-energy ratios for uniform isotropic ($4\pi$) emission. Lighter red and blue lines use diffuse (Lambertian) emission and view-factor (VF) angular weighting. Dark and light gray lines are the corresponding geometry-only photon-interception ratios. The analytical predictions integrate an exponential emitted-energy spectrum over 0.5–12 keV, oblique Kapton attenuation, and normal-incidence Au absorption. The MCNP source begins at 1.001 keV. Geometry-only curves omit Kapton filtering and exceed the deposited-energy ratios near contact. 
}
\label{fig:gap_geometry}
\end{figure}

We used the Monte Carlo radiation transport code MCNP~\cite{Werner2017MCNP62} to quantify how the vacuum gap affects the energy deposited in a 1-$\mu$m-thick Au layer behind 25 or 75~$\mu$m of Kapton, tallied over a 30-$\mu$m-diameter region on the source axis. The source was a uniform 240.8-$\mu$m-diameter disk with isotropic emission and an emitted-energy density proportional to $\exp(-E/kT)$, with $kT = 150$ or 300~eV, sampled between 1.0 and 8.0~keV. Gaps of 1 to 200~$\mu$m were simulated. For each geometry, we divided the Au photon-heating (F6) tally by that from a reference calculation with the same Kapton thickness and $kT$ but a 200~$\mu$m gap, the gap fielded in our experiments. This ratio is independent of the absolute source strength, which can not be well constrained by our measurements. We represent the contact case with a 1~$\mu$m gap to avoid numerical errors associate with a zero-gap geometry in MCNP. Error bars in Fig.~\ref{fig:gap_geometry} show the MCNP statistical uncertainty of the ratio $R$, $\sigma_R = R\sqrt{\epsilon_{\mathrm{near}}^2 + \epsilon_{\mathrm{ref}}^2}$, where $\epsilon$ is the relative error of each tally.

To separate geometric coupling from spectral filtering, we also calculated analytical absorbed-energy ratios for two limiting angular distributions of emission from the source disk: isotropic ($4\pi$) emission and diffuse (Lambertian, labeled VF) emission, the latter corresponding to the standard parallel-disk view factor. Relative to isotropic emission, the diffuse distribution weights each ray by an additional factor of $4\cos\theta$, concentrating emission toward the surface normal. The isotropic case approximates an optically thin emitting layer, and the diffuse case an optically thick emitting surface. For each angular distribution, we computed the fraction of emitted photons intercepted by the 30-$\mu$m Au region and the angular distribution of the intercepted rays. Each ray was attenuated through the Kapton by $\exp[-\mu_{\mathrm{K}}(E)\,t_{\mathrm{K}}/\cos\theta]$, while Au absorption was approximated at normal incidence by $1-\exp[-\mu_{\mathrm{Au}}(E)\,t_{\mathrm{Au}}]$, using tabulated cross sections~\cite{Elam2002}. The resulting response was integrated from 0.5 to 12~keV against an emitted-energy density proportional to $\exp(-E/kT)$ and normalized to the corresponding 200~$\mu$m-gap result. This energy range brackets, with a wide margin, all photon energies that contribute appreciably to heating; photons below 1~keV, which are absent from the MCNP source, are strongly absorbed by the Kapton contribute less than 0.1\% of the modeled heating. We also show geometry-only ratios, which retain only the interception fraction: they omit Kapton attenuation and Au absorption and are therefore independent of $kT$.

At a 1~$\mu$m gap, MCNP gives an enhancement of 2.4--3.4 relative to the 200~$\mu$m gap across the four Kapton-thickness and $kT$ combinations, well below the geometry-only ratios of 5.5--6.7 (isotropic) and 4.3--4.4 (diffuse). As the gap closes, a larger fraction of the intercepted photons arrives at oblique angles and traverses a longer path through the Kapton, so spectral filtering offsets much of the gain in geometric coupling. The offset is strongest for the softer spectrum and, less intuitively, for the thinner Kapton. Seen from the center of the probe at a 1~$\mu$m gap, the edge of the 120.4~$\mu$m-radius source lies $78^\circ$ from the surface normal for 25~$\mu$m of Kapton ($z = 26~\mu$m) but only $58^\circ$ for 75~$\mu$m ($z = 76~\mu$m). With the thinner Kapton, roughly a third of the intercepted photons therefore arrive beyond $60^\circ$, where the Kapton path exceeds twice its thickness; with the thicker Kapton, essentially none do. At 150~eV, the MCNP enhancement is $\sim$36\% of the isotropic geometry-only ratio for 25~$\mu$m of Kapton and $\sim$47\% for 75~$\mu$m; at 300~eV, the corresponding fractions are $\sim$48\% and $\sim$61\%. The isotropic analytical model reproduces the MCNP ratios to within $\sim$8\%, consistent with the isotropic MCNP source, whereas the diffuse model falls 6--22\% lower. The enhancement varies by less than 10\% for gaps below 25~$\mu$m, so thin intermediate layers such as glue bonds have little effect on the deposited energy. For conventional targets with the sample in contact with the ablator, the energy deposited by ablation-plasma x-rays is therefore 2.4--3.4 times that in our 200~$\mu$m-gap experiments, under the assumed spectrum and isotropic emission.

\subsection{Recommendations and Conclusions}

The results of this study clearly indicate that sample heating from absorption of x-rays generated by the ablation plasma is mitigated with uncoated 75-µm Kapton-B ablators. This ablator thickness is also recommended due to the increased levels of spatial smoothing of laser focal spot speckle imprint which can, for thinner Kapton-B layers, result in a wider pressure distribution within the sample \cite{gorman2022}.

Several previous high-pressure studies have employed Al-coated ablators with combined thicknesses of 50-µm or less~\cite{sharma2019,briggs2019,Hodge2022,Madhavi2023,Coleman2022,kraus2021}, with one study reporting the use of a 5-\textmu m-thick CH ablator~\cite{white2020}.~The extent to which this heating mechanism influences interpretation of such experiments depends on both the x-ray absorption of the sample material and the resulting temperature increase. Gold was selected as the primary material in this study because it is highly absorbing to x-rays and has a relatively low specific heat, making it a sensitive indicator of x-ray--induced heating. While the temperature measurements reported here provide a useful benchmark for hydrodynamic simulations used to design high-pressure laser shock experiments, the most direct assessment of sample heating is obtained using the experimental methodology described in this work.

Gold was also the primary target material in two recent laser shock experiments aimed at measuring phase stability and the onset of melting along the principal Hugoniot~\cite{sharma2019,briggs2019}. In those studies, Al-coated 50-\textmu m-thick polyimide ablators and 6.8--8~\textmu m-thick Au foils were employed, using a 351-nm laser wavelength and an average laser irradiance of $I_{\mathrm{Laser}}\sim3\times10^{12}$~W/cm$^{2}$. Based on the results presented here, the presence of an Al coating is expected to produce temperature gradients within the Au samples, with the laser-facing surface reaching temperatures several hundred kelvin above the Hugoniot temperature. Because x-ray diffraction measurements are volume integrated, such gradients could bias the inferred onset pressure of melting to artificially low values. Consistent with this interpretation, Sharma \emph{et al.}~\cite{sharma2019} reported densities of the high-temperature bcc phase that are significantly lower than expected from continuum Hugoniot measurements, suggesting sample temperatures in excess of those along the principal Hugoniot. One approach to mitigating x-ray--induced heating is the incorporation of a thin high-$Z$ heat shield within the target assembly to absorb x-rays before they reach the sample. This strategy is routinely employed on large laser facilities~\cite{wicks2018crystal,coleman2025}.

Experiments employ a range of ablator materials, including Kapton, beryllium, and diamond. Sample heating depends in part on the ablator's x-ray transmission, which is set by its composition and thickness, and on the spectrum the ablation plasma emits. Target designs using beryllium ablators are potentially more susceptible to sample heating because beryllium's low atomic number gives it weaker x-ray attenuation than Kapton or diamond. Additionally, a recent study found that typical impurity concentrations in beryllium ablators may have contributed to substantial sample heating in ramp-compression experiments at higher pressures than those studied here~\cite{Krygier:PRB2026}. At equal thickness, diamond's higher density gives it a larger areal mass than Kapton, reducing transmission of the x rays that heat the sample by an amount that depends on photon energy. Comparisons among ablators must therefore account for both x-ray transmission and material-dependent x-ray emission. The need for, and thickness of, any preheat shielding should likewise be carefully assessed for each application.

The susceptibility of a sample to ablation-plasma x-ray heating depends on how much energy it absorbs, how that energy is distributed, and how much the temperature rises per unit energy. Because the ablator removes the softest photons, the transmitted spectrum is concentrated at a few keV, where high-atomic-number samples such as Ta, Au, Pb, and Bi absorb strongly and absorb energy within a thin surface layer. Heat capacity reinforces this trend: heavier elements have lower specific heats, so the same energy absorbed per unit mass produces a larger temperature rise. Low-atomic-number samples tend to absorb less energy per unit volume and distribute it over a greater depth. Samples thin compared to their absorption length are heated more uniformly, whereas in thick samples the heated region is confined near the surface and contributes less to volume-averaged measurements. How much a given temperature rise matters also depends on the goals of the measurement, so these factors should be evaluated for each target design and experiment.

Experiments also generally vary the laser spot size and target design/geometry. Here we consider a uniform, $4\pi$ isotropic x-ray source arranged as in Fig.~\ref{fig:gap_geometry}(c). Let $R_s$ be the source disk radius, $R_p$ the receiving probe spot radius, and $z$ their plane separation, all in the same units.  For parallel, coaxial disks, the exact geometric interception factor is
\begin{equation}
\Phi(R_s, R_p, z) = \frac{z}{2\pi R_s^2}
  \int_0^{R_s} \rho\,d\rho \int_0^{R_p} r\,dr \int_0^{2\pi} \frac{d\varphi}{R^3},
\end{equation}
where $R^2 = z^2 + \rho^2 + r^2 - 2\rho r\cos\varphi$ is the distance between source and probe area elements [Fig.~\ref{fig:gap_geometry}(c)]. For $R_p\ll R_s$, the geometric interception factor is approximately

\begin{equation}
    \Phi(R_s,R_p,z)\approx
    \frac{R_p^2}{2R_s^2}
    \left(1-\frac{z}{\sqrt{z^2+R_s^2}}\right).
\end{equation}
\noindent This reduced expression is accurate to $\sim$0.4\% in our configuration. We emphasize that this relation only accounts for geometry. As discussed and shown above in Fig.~\ref{fig:gap_geometry}, source spectrum and layer properties can have a significant or even dominant role.

Future work will involve repeating the heating measurements with different gap thicknesses, and larger laser spot diameters. Those measurements would provide experimental constraints to the geometric coupling factor which is needed when trying to relate temperature from the gap-target geometry to the standard no-gap case. In addition, measurements of spectral emission from the laser-generated plasma for a range of laser irradiances would help constrain x-ray source models. Repeat measurements at a drive laser wavelength of 351-nm are needed to quantify x-ray heating levels on other high-power laser facilities \cite{wang2019,guardalben2011,soures1996,Lazicki2021}.

In summary, we experimentally quantify x-ray-induced preheat in laser-driven shock experiments and demonstrate that ablator thickness and Al flash coatings strongly control heating levels. Uncoated 75-µm Kapton-B ablators minimize preheat under MEC-LCLS conditions. Unaccounted x-ray preheat can systematically bias thermodynamic interpretation in laser-driven compression experiments, and we provide the first direct quantitative framework to constrain it experimentally.




\bigskip
\bigskip
\begin{acknowledgments}
We would like to thank Carol Ann Davis for her help in preparing the targets.~This work was performed under the auspices of the US Department of Energy by Lawrence Livermore National Laboratory under contract number DE-AC52-07NA27344 and the Laboratory Directed Research and Development Program at LLNL (project no.'s 26-ERD-023, 27-ERD-013). 
\end{acknowledgments}

\bibliographystyle{elsarticle-num}
\bibliography{MEC}

\clearpage

\onecolumngrid
\noindent
\begin{center}
    \textbf{\color{purple} \Large{SUPPLEMENTARY MATERIALS}}
\end{center}

\begin{center}
    
\bigskip
\Large{\textbf{Quantifying sample heating by ablation-plasma \\x-rays in laser-driven shock experiments}}

\bigskip
\noindent
\normalsize{R.F. Smith, I.K. OCampo, S. Singh, T.M. Hutchinson, T.E. Lockard,
J.H. Eggert, C.P. McGuire, A.L. Coleman, J.M. McNaney, S. Pandolfi,
B. Nagler, D. McGonegle, H.J. Lee, D. Khaghani, N. Czapla, A. Krygier}

\end{center}

\renewcommand{\thefigure}{S\arabic{figure}}

\renewcommand{\thefigure}{S\arabic{figure}}
\renewcommand{\thetable}{S\arabic{table}}
\renewcommand\thesection{S\arabic{section}}
\renewcommand\thesubsection{\thesection.\arabic{subsection}}


\setcounter{table}{0}
\setcounter{figure}{0}
\setcounter{table}{0}
\setcounter{section}{0}
\normalsize 

\bigskip
\bigskip
\section{Diffraction Peak Asymmetry \label{sec:asymmetry}}
\begin{figure}[b!]
\begin{center}
\includegraphics[width=1\columnwidth]{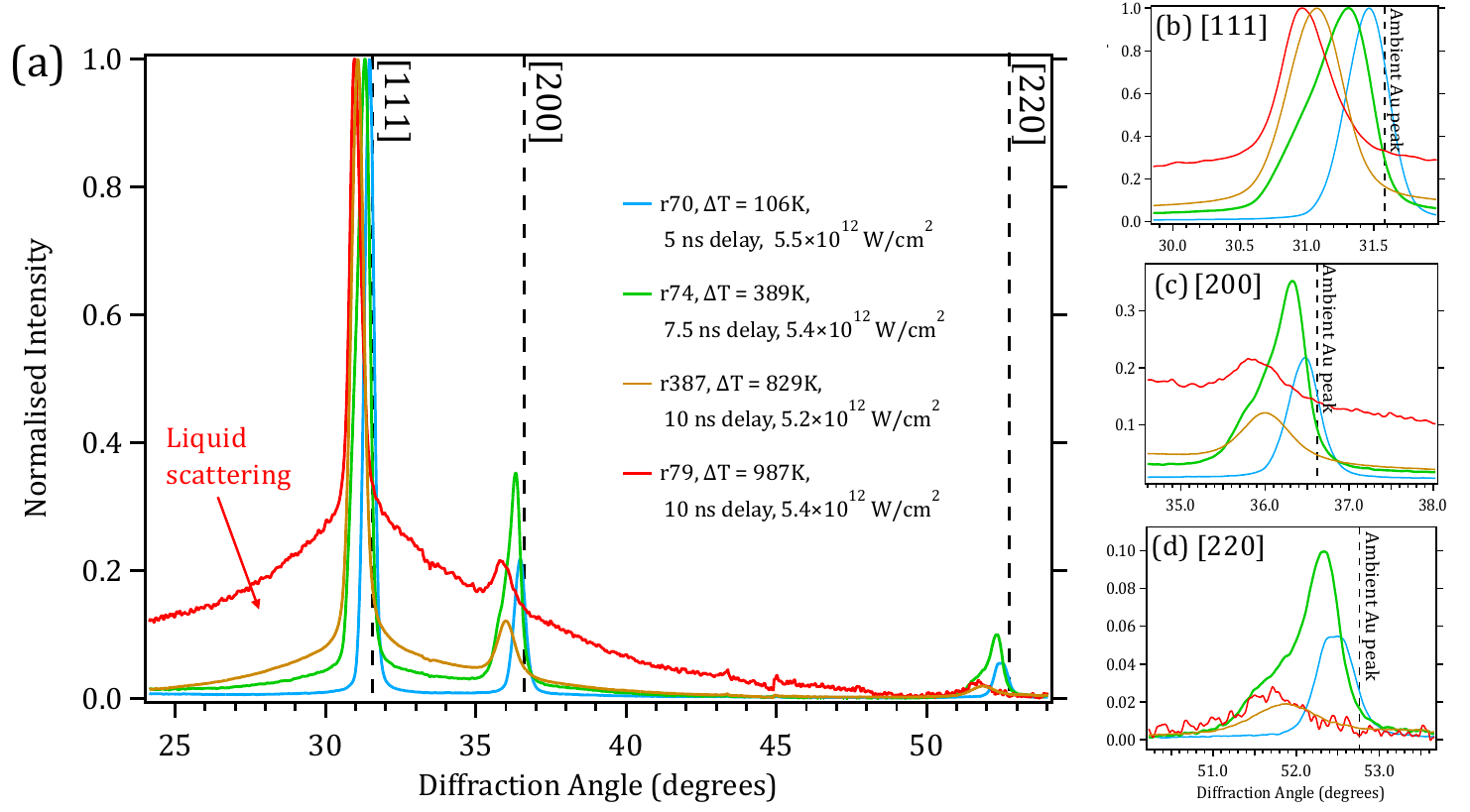}
\end{center}
\caption{\textbf{Diffraction profiles from data shown in Fig. \ref{fig:25um_Kapton}, inset}. (a) Normalized diffraction intensities from a 1-$\mu$m-thick
Au sample as a function of time after the onset of laser illumination, for an average laser irradiance of 5.2$\rightarrow$5.5$\times$10$^{12}$
W/cm$^2$. Each profile represents a separate shot on targets with uncoated 25-$\mu$m-thick Kapton-B ablators. Also shown are the ambient are the ambient peak positions for the Au [111], [200] and [220] peaks (dashed vertical lines). Panels (b), (c) and (d) show a magnified views of the measured peaks. 
}
\label{fig:Peak_Asymmetry}
\end{figure}

\noindent
For some shots we observe an asymmetry in the diffraction peaks which can be explained by depth-dependent x-ray energy deposition and heating. Here, the front of the 1-µm Au layer is hotter than the back, and there is insufficient time for heat flow to fully remove these temperature gradients. As the diffraction signal is volume-integrated the diffraction profiles are a contribution of lattice expansions from these different temperature states. In Fig. \ref{fig:Peak_Asymmetry} we plot the full diffraction profiles for the temperature versus time data from Fig. \ref{fig:25um_Kapton}, inset. For this shot sequence there laser intensity was held approximately constant and the cumulative heating was measured as a function of time. Shots with probe times of 5-ns and 7.5-ns (shot numbers r70 and r74) result in temperatures below melt, whereas shots with probe times of 10-ns have temperatures close to melt and exhibit a mixed solid+liquid scattering.

\begin{figure}[b!]
\begin{center}
\includegraphics[width=1\columnwidth]{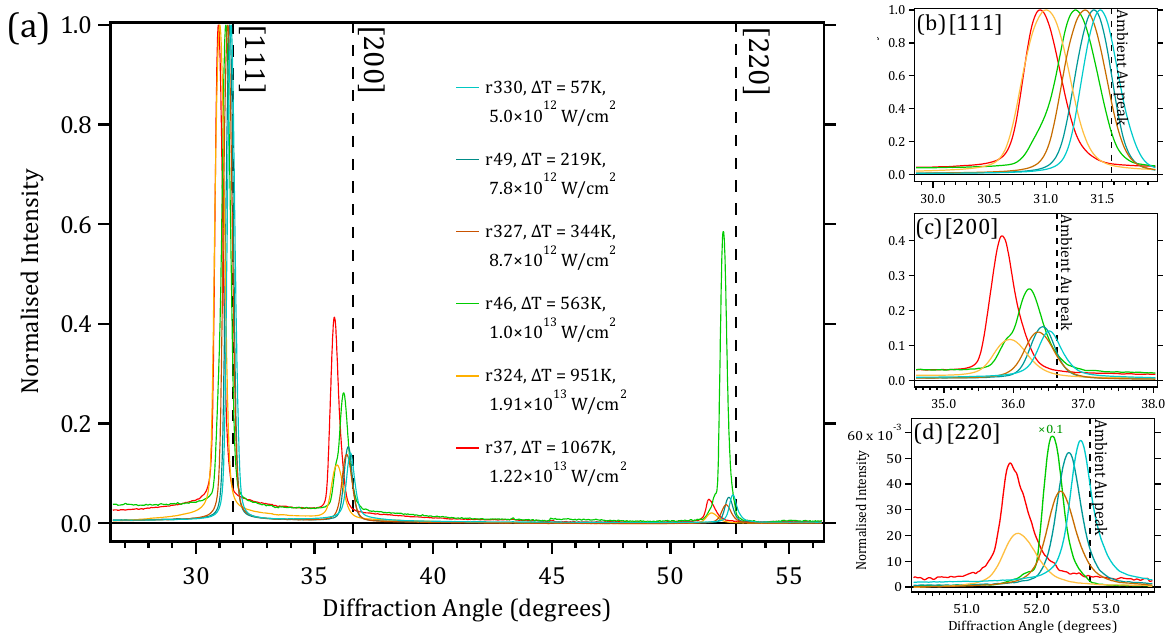}
\end{center}
\caption{\textbf{Diffraction profiles from data shown in Fig. \ref{fig:75um_Kapton} (red-filled circles)}. (a) Normalized diffraction intensities from a 1-$\mu$m-thick
Au sample for a range of laser irradiances. Each profile represents a separate shot on targets with Al-coated 75-$\mu$m-thick Kapton-B ablators, with the x-ray diffraction taken 10-ns after the onset of laser irradiation. Also shown are the ambient are the ambient peak positions for the Au [111], [200], and [220] peaks (dashed vertical lines). Panels (b), (c) and (d) show a magnified views of the measured peaks. 
}
\label{fig:Peak_Asymmetry_75}
\end{figure}

Peak asymmetry is most evident for the higher angle peaks. For the solid-only diffraction the peaks are characterized by a low angle tail. In contrast, for the shots where solid+liquid are observed the asymmetry is characterized by a high angle tail. This difference is due to the shifting functional form of the temperature vs depth profile at the onset of melting. This dependency is also evident for the data shown in Fig. \ref{fig:Peak_Asymmetry_75}, where shot numbers r324 and r37 exhibit evidence for solid+liquid scattering, whereas all the other shots are solid only diffraction.


\end{document}